\documentclass[preprint,showpacs,preprintnumbers,amsmath,amssymb]{revtex4}
\usepackage{graphicx}% Include figure files
\usepackage{dcolumn}% Align table columns on decimal point
\usepackage{bm}% bold math

\newcommand{\ds}{\displaystyle}

\newcommand{\be}{\begin{equation}}
\newcommand{\en}{\end{equation}}
\newcommand{\bea}{\begin{eqnarray}}
\newcommand{\ena}{\end{eqnarray}}
\begin{document}

\title{Tachyonic Universes in Patch Cosmologies with $\Omega>1$}
\author{Sergio del Campo\footnote{Electronic Mail-address:
sdelcamp@ucv.cl}} \affiliation{Instituto de F\'{\i}sica,
Pontificia Universidad Cat\'{o}lica de Valpara\'{\i}so, Avenida
Brasil 2950, Casilla 4059, Valpara\'{\i}so, Chile.}
\author{Ram\'{o}n Herrera\footnote{E-mail address: ramon.herrera@ucv.cl}}
\affiliation{Instituto de F\'{\i}sica, Pontificia Universidad
Cat\'{o}lica de Valpara\'{\i}so, Avenida Brasil 2950, Casilla
4059, Valpara\'{\i}so, Chile.}
\author{Pedro Labra{\~n}a\footnote{E-mail address: plabrana@ubiobio.cl}}
\affiliation{Departamento de F\'{\i}sica, Universidad del
B\'{\i}o–B\'{\i}o, Avenida Collao 1202, Casilla 5-C, Concepci\'on,
Chile.}
\author{Carlos Leiva\footnote{E-mail address: cleivas@uta.cl}}
\affiliation{Departamento de F\'{\i}sica, Universidad de
Tarapac\'a, Casilla 7-D Arica, Chile.}

\author{Joel Saavedra\footnote{E-mail address: joel.saavedra@ucv.cl}}
\affiliation{Instituto de F\'{\i}sica, Pontificia Universidad
Cat\'{o}lica de Valpara\'{\i}so, Avenida Brasil 2950, Casilla
4059, Valpara\'{\i}so, Chile.}
\date{\today}
\begin{abstract}
In this article we study  closed inflationary universe models by
means of a tachyonic field.  We described  a general treatment for
created a universe with $\Omega>1$ in patch cosmology, which is
able to represent General Relativity, Gauss-Bonnet or
Randall-Sundrum patches. We use recent data from astronomical
observations  to constrain the parameters appearing in our model.
\end{abstract}
%
%\PACS{
 %     {PACS-key}{98.80.Bp}   \and
  %    {PACS-key}{98.80.Cq}
   %  } % end of PACS codes
%}

 \maketitle
\section{Introduction}
\label{sec:level1}
 Recent observations from the Wilkinson Microwave Aniso\-tropy Probe
(WMAP)~\cite{wmap,WMAP3, WMAP3a} together with the accurate measurement of
the first acoustic Doppler peak of Cosmic Microwave Background
(CMB)~\cite{Bernardis,benoit, benoit2} are consistent with a
universe having a total energy density very close to its critical
value ($\Omega =1.02\pm 0.04$) \cite{WMAP3, WMAP3a}. Most people interpret
this value as the one corresponding to a flat universe, which is
in agreement with the standard inflationary
prediction~\cite{Guth}. However, this value might also agree with
an alternative point of view of having a marginally open
~\cite{ellis-k} or closed universe ~\cite{Linde:2003hc} with an
inflationary period of expansion at early time.  Indeed, it may be
interesting to consider inflationary universe models in which the
spatial curvature is considered. Therefore, it is interesting to
check if the flatness in the curvature, as well as in the
spectrum, are indeed reliable and robust predictions of
inflation~\cite{Linde:2003hc}.

Nowadays there is a growing interest in the phenomena described by
braneworld scenarios as a mechanism  to localize gravity on them,
achieved through a fine-tuning of the brane tension to the bulk
cosmological constant \cite{RS}. In this way, the brane model
modifies substantially the Friedmann-Robertson-Walker (FRW)
Cosmology.

When brane inflation is considered in the high energy region it is
expected to find important deviations from the standard results in
General Relativity
\cite{bine:yyff,bine2,maart,liddle,ramirez,tsuji,calcagni,kyong}.
Furthermore, as the Gauss-Bonnet (GB) term modifies the Friedmann
equation, it is worth to investigate its effects on the
inflationary processes.
In order to compare these different approaches, we propose a
general formulation of  inflation for a universe dominated by
tachyon matter.

Rolling tachyon matter is associated with unstable D-branes
\cite{Sen:1998sm} and  cosmological implications of this rolling
tachyon were first studied by Gibbons \cite{Gibbons:2002md}.  In
recent years, the possibility of an inflationary phase described
by the potential of a tachyon field has been considered in a quite
diverse topics \cite{Fairbairn:2002yp,Choud, Choud1,Choud2, Choud3, Choud4,Choud5, Choud6, Choud7, Choud8, Choud9, Choud10, Choud11, Choud12, Choud13} and in open and
closed inflationary scenarios in Refs.
\cite{Balart:2007je,Balart:2007gs}.

In this paper we studied closed inflationary universe models where
inflation is driven by a tachyon field. We consider that the
matter content is confined to a four dimensional brane which is
embedded in a five dimensional bulk with a Gauss-Bonnet term.

We follow the approach originally developed by Linde
\cite{Linde:2003hc} for  inflation generated by a standard scalar
field in General Relativity.

%More precisely,  we assume  that a closed universe appears from
%nothing at the point in which $\dot{a} = 0$, $\dot{\phi}=0$, where
%the potential energy density is $V(\phi)$. We start with an
%effective Friedmann equation modified by the Gauss-Bonnet brane
%cosmology with a FRW,  where the  spatial section is closed.

%{\bf We note that the tachyon potential satisfies $dV/d\phi< 0$
%for $\phi
%> \phi_0$ and $V(\phi \rightarrow \infty) \rightarrow 0$. Also, we
%assume that the potential becomes extremely large in the vicinity
%of $\phi<\phi_0$, where the closed inflationary universe appears
%at this point.}

The paper is organized as follows: In Sect. 2 we present the patch
cosmological equations in the tachyon model. In Sect. 3 we
determine the characteristic of a closed inflationary universe
model with a constant potential. In Sect. 4 we study a closed
inflationary scenario with an exponential potentials. We also,
determine the corresponding density perturbations for our model.
Finally,  in Sect. 5 we summarize our results.

\section{Patch Cosmological Equations in  the Tachyon Models}
\label{Sec1}

We start with the five-dimensional bulk action for the
Gauss-Bonnet braneworld:

\begin{eqnarray}
&S& = \frac{1}{2\kappa
_{5}^{2}}\int_{bulk}\!\!d^{5}x\sqrt{-g_{5}}\left\{ R-2\Lambda
_{5}+\alpha \left( R^{\mu \nu \lambda \rho }R_{^{\mu \nu \lambda
\rho }} \right. \right.\nonumber \\ &-&\left. \left.\!\!
4R^{\mu \nu }R_{\nu \mu }+R^{2}\right) \right\}% \nonumber \\
+ \int_{brane}\!\!\!\!\!\!d^{4}x\sqrt{-g_{4}}\left(
\mathcal{L}_{matter}-\sigma \right)\!,\label{action1}
\end{eqnarray}
where $\Lambda _{5}=-3\mu ^{2}\left( 2-4\alpha \mu ^{2}\right) $ is
the cosmological constant in five dimensions, with the $AdS_{5}$
energy scale $\mu=1/l$, $\alpha$ is the GB coupling constant, and
$\kappa _{5}^2=8\pi/m_{5}^3$ is the five dimensional gravitational
coupling constant and $\sigma$ is the brane tension. In the most
standard scenario $\mathcal{L}_{matter}$ describes  a matter content
associated to a scalar field, whose dynamics is determined by the
Klein-Gordon equation. However, in recent models motivated by string
theory, other non-standard scalar field actions have been used in
cosmology. In this context the deep interplay between small-scale
non-perturbative string theory and large-scale brane-world scenarios
has raised the interest in a tachyon field applied to the
inflationary mechanism. We consider that $\mathcal{L}_{matter}$
describes  the dynamics of a tachyon field on the brane.

For a FRW metric, the exact Friedmann-like equation is given by
\cite{Charmousis:2002rc, Charmousis:2002rc1, Charmousis:2002rc2}

\begin{equation}
2\mu \sqrt{1+\frac{H^{2}}{\mu ^{2}}}\left( 3-4\alpha \mu
^{2}+8\alpha H^{2}\right)+\frac{k}{a^{2}} =\kappa _{5}^{2}\left(
\rho +\sigma \right) ,  \label{q3}
\end{equation}
where $\rho$ represents the energy density of the matter sources
on the brane, $a$ is the scale factor, and $k=0,+1,-1$ represents
flat, closed or open spatial section, respectively. Despite the rather complicated form of
Eq. (\ref{q3}), it is possible to make progress if  we use the
dimensionless variable $\chi$ \cite{Lidsey:2003sj},
\begin{equation}
\label{var1} \kappa_5^2(\rho+\sigma) =
\left[{{2(1-4\alpha\mu^2)^3} \over {\alpha} }\right]^{1/2}
\sinh\chi\,,
\end{equation}
The Friedmann equation can be written as
 \begin{equation}
H^2 = {1\over
4\alpha}\left[(1-4\alpha\mu^2)\cosh\left({2\chi\over3}
\right)-1\right]\,,\label{q22}
 \end{equation}
where $\chi$ represent a dimensionless measure of the energy
density. The modified Friedmann equation~(\ref{q22}), together with
Eq.~(\ref{var1}), ensure the existence of one characteristic
Gauus-Bonnet energy scale,
 \be \label{gbscale}
m_\alpha= \left[{{2(1-4\alpha\mu^2)^3} \over {\alpha} \kappa_5^4
}\right]^{1/8}\,,
 \en
such that the GB high energy regime ($\chi\gg1$) occurs if $\rho+\sigma
\gg m_\alpha^4$. Considering the GB term in the action
(\ref{action1}) as a correction to Randall-Sundrum gravity, then
$m_\alpha$ is greater than the Randall-Sundrum energy scale
$m_\sigma=\sigma^{1/4}$, marks the transition to Randall-Sundrum
high-energy corrections to 4D general relativity. Expanding
Eq.~(\ref{q22}) in $\chi$ and using (\ref{var1}), we find in the
full theory three regimes for the dynamical history of the brane
universe,

$\bullet$\,\,Gauss-Bonnet regime (5D),
 \be
\rho\gg m_\alpha^4~ \Rightarrow ~ H^2\approx \left[ {\kappa_5^2
\over 16\alpha}\, \rho \right]^{2/3}\,,\label{gbl}
 \en
$\bullet$\,\, Randall-Sundrum regime (5D),
 \be
 m_\alpha^4 \gg
\rho\gg\sigma \equiv m_\sigma^4 ~ \Rightarrow ~ H^2\approx
{\kappa_4^2 \over 6\sigma}\, \rho^{2}\,,\label{rsl}
 \en
$\bullet$\,\, Einstein-Hilbert regime (4D),
 \be
\rho\ll\sigma~ \Rightarrow ~ H^2\approx {\kappa_4^2 \over 3}\,
\rho\,. \label{ehl}
 \en
Clearly Eqs. (\ref{gbl}), (\ref{rsl}), and (\ref{ehl}) are much simpler than the full
Eq. (\ref{q3}) and in a practical case one of the three energy
regimes will be assumed. Therefore, patch cosmology can be useful
to describe the universe in a region of time and energy in which
\cite{calcagni} \cite{Tsujikawa:2004dm, Tsujikawa:2004dm1, Tsujikawa:2004dm2, Tsujikawa:2004dm3}
\begin{equation}
H^{2}=\kappa_{q}\rho ^{q}-\frac{1}{a^{2}}, \label{friedmann2}
\end{equation}
where $H=\dot{a}/a$ is the Hubble parameter  and $q$ is a patch
parameter that describe the different cosmological model under
consideration. That is, choosing $q=1$ we have the standard
General Relativity with $\kappa_1=8\pi/3m_{p}^2$, where $m_{p}$ is
the four dimensional Planck mass. If we take $q=2$, we have the
high energy limit of brane world cosmology, in which
$\kappa_2=4\pi/3\sigma m_p^2 $. Finally, for $q=2/3$, we have the
GB brane world cosmology, with $\kappa_{2/3}=(\kappa_{5}^2/16\alpha)^{2/3}$, being
$\kappa_5$ the $5D$ gravitational coupling constant and $\alpha=1/8g_s$
is the GB coupling ($g_s$ is the string energy scale). In brane
world cosmology in five dimensions the matter is confined to a
four-dimensional brane, while gravity can propagate into the bulk.
On the other hand, the energy conservation equation  on the brane
follows directly form the Gauss-Codazzi equations and, it is
reduce to the usual form,
\begin{equation}
\dot{\rho}+3H\left( \rho +P\right) =0,  \label{q4}
\end{equation}
where $\rho$ and $P$ represents the energy and the pressure
densities, respectively.

When $\mathcal{L}_{matter}$ described the dynamics of the tachyon
field, the expression for $\rho$ and $P$ are \cite{Sen:1998sm}

 \be
\rho = \frac{V(\phi)}{\sqrt{1-\dot{\phi}^2}}\label{den}, \en and
\be P = -V(\phi) \sqrt{1-\dot{\phi}^2}\label{pre}, \en
respectively, where $V(\phi)$ is the scalar tachyonic potential.
The energy conservation (\ref{q4}) can be written as
 \be \ds
\frac{\ddot{\phi}}{1-\dot{\phi}^2}\,=\,-3\,\frac{\dot{a}}{a}\,\dot{\phi}\,-\,
\frac{1}{V(\phi)}\, \frac{d V(\phi)}{d\phi}\,\,\label{ec3}. \en

On the other hand, from the effective Friedmann equation
(\ref{friedmann2}) we obtain the equation of motion for the scale
factor, \be \ddot{a}=a\kappa_q
\rho^{q-1}\left[\rho\left(1-\frac{3}{2}q\right)-\frac{3}{2}\,q\,P\right],
\label{addot} \en where the dot  denotes derivative with respect
to the time $t$. For convenience we will use  units in which
$c=\hbar=1$.

\section{Constant Potential}
\label{sec1}

Following the treatment developed in Refs.~\cite{Linde:2003hc} and
\cite{Balart:2007gs}, we can study a closed inflationary universe,
where inflation is driven by the tachyon field.  Let  us start by
consider a simple  model with the following step-like effective
tachyon potential: $V(\phi)=V_0 = constant$ for $\phi > \phi_0$,
and $V(\phi)$ is extremely steep for $\phi <\phi_0$. We consider
the birth of an inflating closed universe which can be created
"from nothing" in a state where the tachyon field takes the value
$\phi_{in} \leq \phi_0$ at the point in which $\dot{a}=0$,
$\dot{\phi}=0$. The potential energy density in this point is
$V(\phi_{in})$ . We consider a first phase where the tachyon field
instantly falls down to the value $\phi_0$. Since this process
happens nearly instantly we can consider $\dot{a}=0$, so that the
tachyon field arrives  to the end of this first stage with a
velocity given by:

 \be \ds
\dot{\phi}_0\,=\,
\sqrt{1-\Big(\frac{V(\phi_0)}{V(\phi_{in})}\Big)^2}
\,\,\label{ec2b}, \en where we have considered the positive sign
of the square root because the tachyon field increase during this
process.

 We can see from Eq.(\ref{addot}), that the condition $\ddot{a}=0$ is satisfied if \be
\rho\left(1-\frac{3}{2}q\right)=\frac{3}{2}q\,P.\en With the
suitable replacing, this means that we have three different
scenarios, related to the following conditions:

\begin{eqnarray}
\ddot{a}&=&0 \Leftrightarrow \dot{\phi}_0^2=\frac{2}{3q},  \\
\ddot{a}&<&0 \Leftrightarrow \dot{\phi}_0^2>\frac{2}{3q},
\\\ddot{a}&>&0 \Leftrightarrow \dot{\phi}_0^2<\frac{2}{3q} ,
\end{eqnarray}
where the first condition implies, if initially $\dot{a}=0$, then
the universe remain static and the tachyon field moves with constant
speed given by Eq.~(\ref{ec2b}). In the second condition the
universe starts to  move  with negative acceleration ($\ddot{a}<0$)
from the state $\dot{a}=0$ and  the tachyon field equation
describing negative friction, so that the $\phi$ field moves faster,
and thus $\ddot{a}$ becomes more negative. This universe rapidly
collapses. In the last condition, we have $\ddot{a}>0$, and   the
universe enters into an inflationary stage.

From now on, we will consider  the patch cosmological equations of
motion (\ref{ec3}) and (\ref{addot}) in the cases where the
condition $\ddot{a}>0$ is satisfied. Note that, in the regimen
where $V(\phi_0)=V_0=const.$, the solution of the scalar field
equation (\ref{ec3}), results in:

\be \dot{\phi}^2=\frac{1}{1+Ca^6},\label{phidotsq}\en where
 $C(1-\dot{\phi}^2_0)/(\dot{\phi}^2_0a_0^6)$. Due to this, the
evolution of the universe rapidly falls into an exponential
regimen (inflationary stage) where the scale factor becomes $a\sim
e^{H_0\,t}$, with  Hubble parameter given by
$H_0=\sqrt{\kappa_q\,V_0^q}$. We can now integrate
Eq.(\ref{phidotsq}) and obtains

\be \triangle
\phi_{inf}=\frac{1}{3\sqrt{\kappa_q\,V_0^q}}
\ln\left[\frac{1}{\sqrt{C}}+\sqrt{1+\frac{1}{C}}\right].\label{triangle}
\en This means that when the universe enters in to the
inflationary stage, the tachyon field moves by the amounts
$\triangle \phi_{inf}$ and then stop.

At early time, before inflation takes place, we can write
conveniently the equation for the scale factor as follows: \be
\ddot{a}(t) = 2\kappa_1\,V_0\,a(t)\,\beta(t) \label{beta1}. \en
 Here, we have introduced a small
 time-dependent dimensionless  parameter $\beta (t)$:
  \be \beta(t)
 =\frac{1}{2}\frac{\kappa_q}{\kappa_1}\frac{\rho^{q-1}}{V_0}
 \left[\rho\left(1+\frac{3}{2}q\right)-\frac{3}{2}q\,P\right],
 \label{def-bet} \en
 or, by using  suitable variables in terms of the tachyon field,
 we obtain

\be \beta(t)=\frac{1}{2}\frac{\kappa_q}{\kappa_1}\frac{V_0^{q-1}}
{(1-\dot{\phi}^2)^{q/2}}\left[1-\frac{3}{2}q\,\dot{\phi}^2\right]
 \label{def2-bet}, \en
note that $\beta(t)\rightarrow 0$ when $\dot{\phi}^2(t)\rightarrow
2/3q$.

Now we proceed to make an   analysis for the model in the case
$\beta(0)\equiv\beta_0 \ll 1$. At the beginning of the process we
have $a(t) \approx a_0$ and $\beta(t) \approx \beta_0$. Then,
Eq.(\ref{beta1}) takes the form: \be \ddot{a}(t) 2\kappa_1\,a_0\,\beta_0, \en and thus
 \be \ds
a(t) = a_0 \left(1 + \kappa_0 \,\beta_0\,V_0\,t ^2\right).
\label{adet} \en

From Eqs. (\ref{phidotsq}) and (\ref{adet}) we find that at a time
interval where $\beta(t)$ becomes twice as large as $\beta_0$,
$\Delta t_1$ is given by

\be \Delta
t_1=\left[\frac{\left(1-\dot{\phi}^2_0\right)^{q/2}}{3q\,\kappa_q\,
\dot{\phi}^2_0\left(1+\frac{3}{4}(q-2)\dot{\phi}^2_0\right)}\right]^{1/2}
\label{tiempo}, \en consequently the tachyonic field increases  by
the amount \be \ds \Delta \phi_1 \sim  \dot{\phi}_0 \, \Delta t_1
\sim \frac{1}{\sqrt{3q\,\kappa_q\,V_0^q}}\,\,\label{bexp}, \en where
we have kept only the first term in the expansion of $\Delta t_1$.
Note that this result depends on the values of the patch parameter
$q$ and on $V_0$ and the increase of the tachyonic field is less
restrictive than the standard scalar field, in which $\Delta \phi_1
=const.\sim -1/(2\sqrt{3\pi})$ \cite{Linde:2003hc}. After the time
$\Delta t_2 \approx \Delta t_1$, where now the tachyonic field
increases by the amount $\Delta\phi_2 \approx \Delta\phi_1$, the
growth rate  of $a(t)$ also increases. This process finishes when
$\beta(t) \rightarrow \beta_f$, where
$\beta_f=\kappa_q\,V_0^{q-1}/(2\kappa_1)$. Since, at each interval
$\Delta t_i$ the value of $\beta$ doubles, the number of intervals
$n_{int}$ after which $\beta(t) \rightarrow \beta_f$ is
 \be \ds
n_{int} =\frac{\ln \beta_f-\ln \beta_0}{\ln 2} \,\,\label{n}. \en

Therefore, if we know the initial velocity of the tachyon, we can
estimate the value of the tachyon field at which  inflation begins
by means of the expression

\be \ds \label{ficero} \phi_{inf} \sim \, \phi_0  +\frac{1 }{\ln
2}\;\left(\ln \beta_f - \ln \beta_0 \right)
\!\frac{1}{\sqrt{3q\,\kappa_q\,V_0^q}} \,\,. \en This expression
indicates that our result for $\phi_{inf}$ is sensitive to the
choice of particular value of the patch parameter and of the
potential energy $V_0$, apart from the initial velocity of the
tachyonic field $\phi$ immediately after it rolls down to the
plateau of the potential energy. Note that in the special case
where  $q=1$ we reproduce the previous  results for $\phi_{inf}$
obtained in  Ref.\cite{Balart:2007gs}.

Now let us study the implications of these results for the theory
of quantum creation of a closed inflationary universe in this
model. The probability of the creation of a closed universe from
nothing was computed in Ref. \cite{Koya}, and can be written as

\be P\sim \exp\left[-\beta_0\,f\left(V_0,\kappa_q\right)\right],
\en where $f(V_0,\kappa_q)$ is a function of the $V_0$ and the
parameter $\kappa_q$. This expression tells us that the
probability of creation of the universe with $\beta_0 \neq 0$ is
not exponentially suppressed if $\beta_0 < \frac{V_0}{M^{4}_{p}}$,
$\beta_0 < \frac{V_0^4}{\sigma^3\,M^{4}_{p}}$, and $\beta_0 <
M_p^2\,\kappa_{2/3}^3/(2\pi^2)$ for $q=1$, $q=2$ and $q=2/3$,
respectively.

\section{Exponential Potential in patch cosmologies}
\label{sec2}

\begin{figure}[th]
\includegraphics[width=4.0in,angle=0,clip=true]{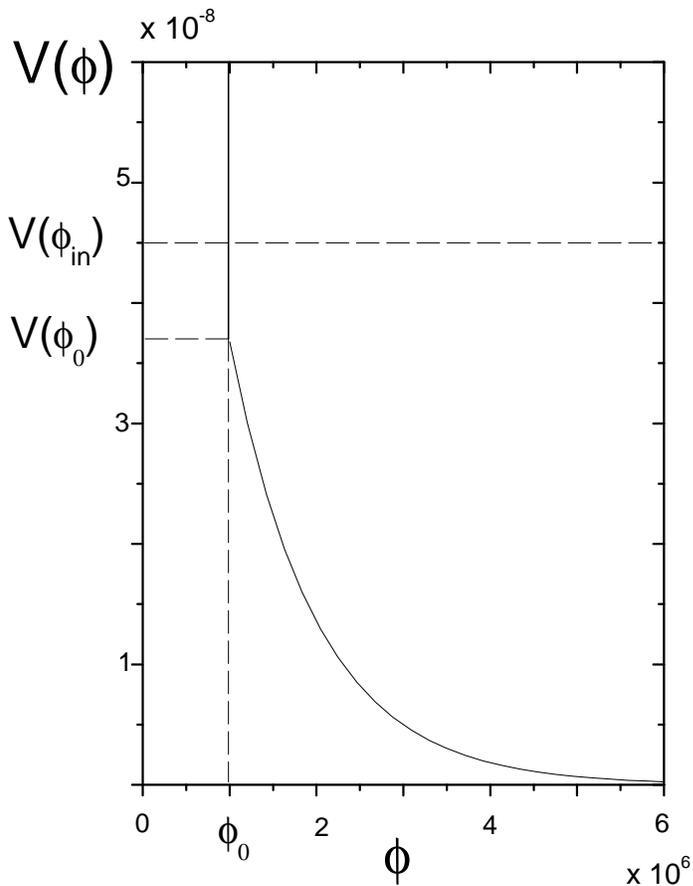}
\caption{ Tachyonic potential as a function of the tachyon field
$\phi$. The values of the parameters were chosen as
$A=10^{-7}\kappa^{-2}$
 and $\lambda=10^{-5}\kappa^{-1/2}$, in  units
$\kappa=8\pi\,G=8\pi/M_p^2=1$.} \label{pot1}
\end{figure}

In order to consider a more standard  tachyonic model  we are
going to study a patch cosmologies in which the effective
potential is given by

\be \label{pot} V(\phi) \simeq A\,e^{-\lambda \phi}, \en here
$\lambda$ ($\lambda> 0$, in units $M_p$) is related to the tachyon
mass \cite{Fairbairn:2002yp}, and $A$ represents a free parameter,
that can be tuned by the observations. We will also assume that the
effective potential sharply rises to indefinitely large values in
the vicinity of $\phi=\phi_0$, see Fig.\ref{pot1}.
%%%%%%%%%%%%%%%%%%%%%%%%%%%%%%%%%%%%%%%%%%%%%%%%%%%%%%%%%%%%%

%%%%%%%%%%%%%%%%%%%%%%%%%%%%%%%%%%%%%%%%%%%%%%%%%%%%%%%%%%%%%%%%%%

The whole process is composed  by different stages. The first stage
corresponds to the creation of the closed universe ``from nothing"
in a state where the tachyon field takes the value $\phi_{in} \leq
\phi_0$ at the point in which $\dot{a}=0$, $\dot{\phi}=0$,  and
where the potential energy is $V(\phi_{in})$ \cite{Balart:2007gs}.
If the effective potential for $\phi\ < \phi_0$ grows very sharply,
then the tachyon field instantly falls down to the value $\phi_0$,
with potential energy $V(\phi_0)$, and the initial potential energy
becomes converted to  kinetic energy. Then, we find $
\dot{\phi}^2_0= 1- \left(\frac{V(\phi_0)}{V(\phi_{in})}\right)^2$.

Following the discussion of the previous section we assume that the
initial condition $\dot{\phi}^2<\frac{2}{3q}$ is satisfied. In terms
of the potential this condition reads

\be \label{cond} \sqrt{\frac{3q-2}{3q}} <
\frac{V(\phi_0)}{V(\phi_{in})} < 1, \en which ensures  that the
patch cosmology enters to an inflationary regimen. As it was
mentioned previously, in all other cases the universe either remains
static or it collapses rapidly.

The other steps are described by Eqs.(\ref{ec3}) and (\ref{addot})
in the interval $\phi \geq \phi_0$ with initial conditions
$\dot{\phi}=\dot{\phi}_0$, $a=a_0$ and $\dot{a}=0$. In particular,
the second part of the process corresponds to the motion of the
tachyon field before the beginning of the  inflation stage, and it
is well described by the following approximated  field equations
of motion:

\begin{eqnarray}
\frac{\ddot{\phi}}{1-\dot{\phi}^2} -3\,\frac{\dot{a}}{a}\,\dot{\phi}\,,
\label{t1}\\
\nonumber \\
\label{t2} \ddot{a} = 2\kappa_1\,a\,V(\phi)\,\beta(t),
\end{eqnarray}
where  $\beta(t)$ satisfies the constraint  $\beta(t)\ll 1$, as
before.

The last statement corresponds to the stage of inflation where
$\dot{\phi}$ is small enough and the scale factor $a(t)$ grows  up
exponentially. This period is well described by the following
equations of motion \cite{Sami:2002fs,Paul:2003jx}:

\be \ds 3\,\frac{\dot{a}}{a}\,\dot{\phi} = -\frac{1}{V}\,
\frac{dV}{d\phi}\,\,\label{inf1}, \en
\begin{equation}
\label{inf2} \ddot{a} = \kappa_q\,a\,V(\phi) ^q\,.
\end{equation}

Let us consider the second stage in a more detailed way. After the
scheme of section \ref{sec1} we can solve the equation for $a(t)$
by considering $\beta(t) \ll 1$. Then, at the beginning of the
process, when $a\approx a_0$ and $\beta \approx \beta_0$, the
increment of the tachyon field during the time $\Delta t$, which
makes the value of $\beta$ twice as great as $\beta_0$, is $\Delta
\phi \approx (3\,q\,\kappa_q\, V(\phi_0)^q)^{-1/2}$. This process
continues until $\dot{\phi}$ is small enough so that the universe
begins to expand in an exponential way, which characterizes the
inflationary era. We assume that inflation begins when $\beta(t)$
approaches to $\beta_f$, where $\beta_f=
\kappa_q\,V(\phi_0)^{q-1}/(2\kappa_1)$. Then, according to our
previous result,  the tachyon field gets the value

 \be
 \ds \phi_{inf} \sim \, \phi_0  +\frac{1}{\ln 2}\; \!\left(\ln \beta_f -\ln
\beta_0\right)\; \frac{1}{\sqrt{3\,q\,\kappa_q\, V(\phi_0)^q}}.
\label{eq20} \en

During inflation, the scalar factor is given by
\begin{equation}
\frac{a}{a_o}=\exp\left(-3\kappa_q\,\int_{\phi_o}^{\phi}\,\frac{V^{q+1}}{dV/d\phi^{'}}\,d\phi^{'}\right),
\end{equation}
and the corresponding $N$ e-folds results to be:
\begin{equation}
N=\frac{1}{2}\,\left[\left(\frac{V(\phi_{inf})}{V_f}\right)^q-1\right],\label{NN}
\end{equation}
where $V_f=V(\phi_f)=(q\,\lambda^2/(6\kappa_q))^{1/q}$ is the
value of the tachyon potential at the end of inflation
\cite{Paul:2003jx}.

Then, by using Eqs.(\ref{eq20}) and (\ref{NN})  we can relate the
value of $\beta_0$ with the number of e-folds, $N$:

\begin{equation}
\beta_0=\left(\frac{\kappa_q\,V(\phi_0)^{q-1}}{2\kappa_1}\right)\;
\left[\frac{q\,\lambda^2\,(2N+1)}{6\kappa_q\,V(\phi_0)^q}\right]^{\left(\frac{\ln
2}{q\,\lambda}\right)\sqrt{3q\kappa_q\,V(\phi_0)}} .\label{ff}
\end{equation}
Note that, in the  case $q=1$, we recover the previous  result
obtained in Ref.\cite{Balart:2007gs}.

Let us assume for definiteness  that for $N=60$ one  would  have
$\Omega=1.1$. Then one can show that for $N=59.5$ and the same
value of the Hubble constant one would have $\Omega\approx1.3$
\cite{Linde:2003hc,Balart:2007gs}.
 In order to have the value of $\Omega$ in the range $1 \leq \Omega< 1.1$
 we require  a fine tuning of the value of
$V(\phi_0)$.

Now, we are going to compute the value of the scalar field
$\phi_0$. For this, we consider the mass of the tachyon to be
$\lambda=10^{-5}\kappa^{-1/2}$ and the number of e-folds $N=60$.
Then, for the case $q=1$ and if we assume $\phi_0\sim
10^{5}\kappa^{1/2}$, then in order to satisfy
$\beta_0<V(\phi_0)/M_p^4$ we have $\beta_0<2\times 10^{-10}$.
Following  Refs.\cite{Linde:2003hc} and \cite{Balart:2007gs} we
see  that the probability to begin with the value
$\beta_0\ll\,2\times 10^{-10}$ is suppressed due to the small
phase space corresponding to this value of $\beta_0$. Thus, it is
most probable to have $\beta_0\sim 2\times 10^{-10}$, and in this
case, if we set $\phi_0=1.1\times10^{5}\kappa^{1/2}$, which
satisfies the condition $\beta_0<V(\phi_0)/M_p^4$, we obtain
$N=60$ which correspond to $\Omega=1.1$. On the other hand, if we take $\phi_0 = 0.5 \times 10^5\,\kappa^{1/2}$, we get
N = 171 and the universe becomes flat. In the case $q=2$,
following the same argument as before, we have  $\beta_0 \sim
10^{-12}$ then, for $N=60$ which correspond to $\Omega=1.1$, we set
$\phi_0\simeq3\times10^{5}\kappa^{1/2}$. For $q=2/3$ we have
$\beta_0 \sim 10^{-11}$ and $\phi_0\sim
1.5\times10^{5}\kappa^{1/2}$. We would like to note that the only way to obtain inflationary universe with
 $\Omega> 1$ is to assume that the universe inflated only by
about $e^{60}$ times. In order to explain this point we assume that
for $N = 60$ one would have  $\Omega= 1.1$ for all values of patch parameter $q$. Then, it is possible to
show that for N = 59.5 one would have $\Omega= 1.3$, whereas for
$N = 60.5$ one would have $\Omega= 1.03$ corresponding to a marginally closed universe. Then, for $N>60$ (i.e $\phi_0<1.1\times10^{5}\kappa^{1/2}$, which q=1, $\phi_0<3\times10^{5}\kappa^{1/2}$ which $q=2$ and $\phi_0<
1.5\times10^{5}\kappa^{1/2}$ for $q=2/3$) the universe becomes flat.
Thus in order to
obtain $\Omega= 1.1\pm 0.05$ one would need to have $N = 60$
with accuracy of about 1 \cite{Linde:2003hc}.

\section{Perturbations }
\label{Sec5}

The consequences of the dynamics of a closed inflationary universe,
such as, slow roll parameters, density perturbations and tensor
perturbations are quite complicate to be realized, since it involves
several contributions.
As it was shown in Refs.
\cite{Linde:2003hc,Balart:2007je,delCampo:2004gh,del Campo:2004ee, del Campo:2004ee1,ramon}, in a closed inflationary universe model this
correction should be somewhat modified during the very first
stages of the inflationary period, at $N = {\cal O}(1)$. For
instance, if we consider $N \geq 60$ in order to solve the
cosmological "puzzles", we may get rid of the corrections and
consider the standard flat-space expressions which gives correct
results for $N> 3$. In particular, the amplitude of scalar
perturbations for a flat space, generated during tachyon inflation
is defined in Ref. \cite{Hwang:2002fp}
 \be P_{\cal
R}=\left(\frac{H^2}{2\pi\dot{\phi}}\right)^2\;\frac{1}{V}.
\label{ec10}
 \en

%If one interprets perturbations produced immediately after the
%creation of closed universe (at $N\sim O(1)$) as perturbations on
%the horizon scale $l\sim 10^{28} cm$, then the maximum at $N\sim
%10$ would correspond to the scale $l\sim 10^{24} cm$, and the
%maximum at $N\sim 15$ would correspond to the scale $l\sim 10^{22}
%cm$, which is similar to the galaxy scale.

One interesting parameter to consider is the so-called spectral
index $n_s$, which is related to  $P^{1/2}_{\cal R}(k)$
\cite{Hwang:2002fp}. For modes with a wavelength much larger than
the horizon ($k \ll a H$), the spectral index $n_s$ is an exact
power law, expressed by $P^{1/2}_{\cal R}(k) \propto k^{n_s-1}$,
where $k$ is the comoving wave number. From WMAP five-year data it
is obtained the values $P_{\cal R}\simeq 2.4\times\,10^{-9}$ and
the spectral index $n_s\simeq0.96$ ($\Lambda $CDM model)
\cite{WMAP3, WMAP3a}.
%%%%%%%%%%%%%%%%%%%%%%%%%%%%%%%%%%%%%%%%%%%%%%%%%%%%%%%%%%%%%
 \begin{figure}[th]
\includegraphics[width=4.0in,angle=0,clip=true]{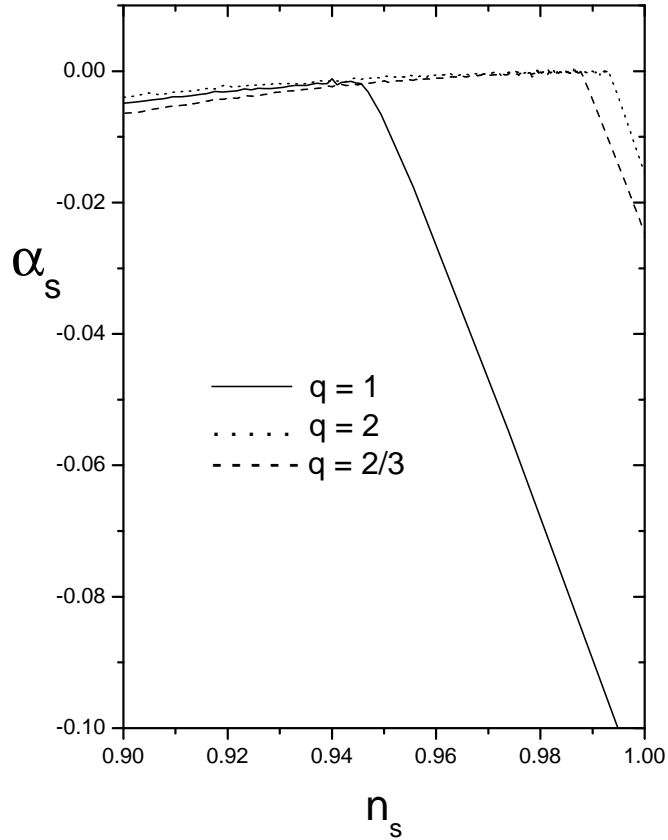}
 \caption{The plot shows the running of the scalar spectral index
$\alpha_s$  as a function of the $n_s$. We have taken $\sigma=10^{-10}$,
$\kappa_{2/3}=10^{-3}$ and we have used unit where $m_p=1$. } \label{fig2}
\end{figure}
%%%%%%%%%%%%%%%%%%%%%%%%%%%%%%%%%%%%%%%%%%%%%%%%%%%%%%%%%%%%%%%%%%
In tachyon inflationary models the scalar spectral index $n_s$ and
the tensor spectral index $n_T$ are given by
\begin{equation}
n_s=1-2\,(2\epsilon_1 - \epsilon_2+\epsilon_3)\label{ns},
\end{equation}
and \be n_T = -2(\epsilon_1-\epsilon_3),\en
 in the slow-roll
approximation \cite{Hwang:2002fp, Hwang:2002fp1, Hwang:2002fp2}. Here, the slow-roll parameters
are defined by:
\begin{equation}
 \epsilon_1 \simeq-\frac{\dot{H}}{H^2}\,,\;\;
\epsilon_2 \simeq
-\frac{\ddot{\phi}}{H\,\dot{\phi}}\;,\;\epsilon_3\simeq\;-
\left(\frac{\dot{\phi}}{2\,H\,V}\right)\,\frac{dV}{d\phi}.\label{e2}
 \end{equation}

One of the  features of the five-year data set from WMAP is that
it suggests a significant running in the scalar spectral index
$dn_s/d\ln k=\alpha_s$ \cite{WMAP3,WMAP3a}. From Eq.(\ref{ns}) we obtain
that the running of the scalar spectral index for our model
becomes

\begin{equation}
\alpha_s=\frac{d n_s}{d\ln
k}\simeq\;\left(\frac{2V_{,\;\phi}}{3\kappa_q\,V^{q+1}}\right)\,
[2\;\epsilon_{1\;,\;\phi}-\epsilon_{2\;,\;\phi}+\epsilon_{3\;,\;\phi}],\label{dnsdk}
\end{equation}
where we have used that $d\ln k=-dN$.

In models with only scalar fluctuations, the marginalized value
for the derivative of the spectral index is approximated to
$dn_s/d\ln k=\alpha_s \sim -0.03$ for WMAP five-year data only
\cite{WMAP3,WMAP3a}.

In Fig.2 we plot the running spectral index $\alpha_s$ versus the
scalar index $n_s$ . In doing this, we have taken three different
values of the patch parameter $q$, for the same initial condition.
We note that for $n_s=0.97$, we obtain $\alpha_s\simeq -0.0482$,
$\alpha_s\simeq -0.0003$ and $\alpha_s\simeq -0.0005$ for $q=1$,
$q=2$, and $q=2/3$, respectively.
\section{Conclusion and Final Remarks}
\label{Sec6}

In this paper we have studied a single tachyonic field  in closed
inflationary universe models in which the gravitational effects
are described by patch cosmology. In this theory, the Friedmann
equation has a complicated structure shown by Eq. (\ref{q3}) and
in the limits of Gauss-Bonnet, high energy Randall Sundrum II
model and 4D regime we obtain an effective Friedmann equation
whose form is given by: $H^2 \sim \rho^q$, where $q=2/3$, $q=2$
and $q=1$ for the GB, high energy brane world cosmology and 4D
Einstein's Theory of Relativity, respectively. We studied the
patch cosmologies for two different models, corresponding to a
constant potential and to a self-interacting tachyonic potential
given by $V=V_0\exp(-\lambda\,\phi)$, ($\lambda>0$). In the first
scenario, we have considered a potential with  two regimes, one
where the potential is constant and another one where the
effective potential sharply rises to infinity. In the context of
Einstein's theory of GR, this model was studied by Linde
\cite{Linde:2003hc} with one standard scalar field. He showed that
this model was not satisfactory because  a constant  potential
implies that the universe collapses too soon or inflates forever.
The tachyonic case was studied in Ref. \cite{Balart:2007gs}  in
this model it was shown that it is less restrictive than the
former. In our cases, we can fix the graceful exit problem because
in patch cosmologies  we have some extra ingredients related to
the value of the patch parameter $q$ and the patch gravitational
constant $\kappa_q$, that allows to reach the value
$\dot{\phi}^2=2/3q$, which is needed to terminate inflation, see
Eq.(\ref{triangle}). The inclusion of corrections to the
gravitational theory changes some of the characteristic of the
perturbations. For instance, in patch cosmologies  we found that
the characteristic of the running spectral index $\alpha_s=0$
(standard theory \cite{pp1}) changes to $\alpha_s\neq 0$ by virtue
of equation (\ref{ns}). In particular for $n_s=0.97$, we obtain
$\alpha_s\simeq -0.0482$, $\alpha_s\simeq -0.0003$ and
$\alpha_s\simeq -0.0005$ for $q=1$, $q=2$, and $q=2/3$,
respectively. These values are  not far from the values given by
WMAP five-year data  \cite{WMAP3,WMAP3a}.

Summarizing, we have been successful in describing a closed
inflationary universe in a patch cosmology with a tachyon field
theory.
\begin{acknowledgments}
This work was supported by Grants FONDECYT
1070306 (SdC), 1090613 (RH) and 11060515 (JS). Also it
was partially supported by PUCV DI-PUCV 2009. P. L. is supported by Direcci\'on de Investigaci\'on de la Universidad del B\'{\i}o-B\'{\i}o through Grant No. 096207 1/R  . C.L was supported by Grant UTA DIPOG 2009-2010. R.H. and J. S. wish to thank Departamento de
F\'{\i}sica de la Universidad de Tarapac\'a de Arica for its kind
hospitality.
\end{acknowledgments}
%
% BibTeX users please use
% \bibliographystyle{}
% \bibliography{}
%
% Non-BibTeX users please use

\end{document}